# Binary square axicon with chiral focusing properties for optical trapping

**B. VINOTH,**[1,5] **A. VIJAYAKUMAR,**[2,5] **MANI RATNAM RAI,**[2] **JOSEPH ROSEN,**[2,*] **CHAU-JERN CHENG,**[1] **OLEG V. MININ**[3,4] **AND IGOR V. MININ**[4]

[1]*Institute of Electro-Optical Science and Technology, National Taiwan Normal University, Taipei 11677, Taiwan, ROC.*
[2]*Department of Electrical and Computer Engineering, Ben-Gurion University of the Negev, P.O. Box 653, Beer-Sheva 8410501, Israel.*
[3]*Tomsk State University, 36 Lenina Avenue, 634050 Tomsk, Russia.*
[4]*Tomsk Politechnic University, 30 Lenina Avenue, 634050 Tomsk, Russia.*
[5]*The authors contributed equally to the work.*
*\*rosenj@bgu.ac.il*

**Abstract:** We introduce a novel phase-only diffractive optical element called chiral binary square axicon (CBSA). The CBSA is designed by linearly rotating the square half-period zones of the binary square axicon with respect to one another. A quadratic phase mask (QPM) is combined with the CBSA using modulo-$2\pi$ phase addition technique to bring the far-field intensity pattern of CBSA at the focal plane of the QPM and to introduce quasi-achromatic effects. The periodically rotated zones of CBSA produces a whirlpool phase profile and twisted intensity patterns at the focal plane of QPM. The degree of twisting seen in the intensity patterns is dependent upon the angular step size of rotation of the zones. The intensity pattern was found to rotate around the optical axis along the direction of propagation. The phase patterns of CBSA with different angles of zone rotation are displayed on a phase-only spatial light modulator and the experimental results were found to match with the simulation results. To evaluate the optical trapping capabilities of CBSA, an optical trapping experiment was carried out and the optical fields generated by CBSA were used for trapping and rotating yeast cells.

## 1. Introduction

Diffractive optical elements (DOEs), in general, can modulate the optical fields to generate any intensity and phase distributions [1-3]. Unlike refractive optical components [4, 5], DOEs are compact, lightweight and thinner and can be easily integrated to any optical system [6, 7]. With development in fabrication technology, the DOEs are manufactured with a high resolution, multiple phase levels and have demonstrated a high efficiency in par with the refractive counterparts [6]. Due to the aforementioned facts, DOEs have been used for tailoring optical beams for various applications such as imaging [8], eye correction [9], encryption and security [10], biomedical and industrial applications [9, 11]. Besides the above applications, the special intensity and phase profiles generated by the DOEs are used extensively for optical trapping [11].

Optical trapping is a technique to trap micro objects in a confined area in space using the forces of light [12, 13]. The idea of trapping particles using light was introduced as early as 1960s [14, 15]. The first optical traps used a tightly focused Gaussian beam with a waist diameter comparable to the size of the object. Such optical traps function using the gradient force field of the Gaussian intensity profile [16]. Even though this conventional optical trap is capable of trapping, moving and isolating atoms and microorganisms there are several drawbacks [17, 18] accompanying it. In the Gaussian beam trap, the trapped sample is overlapped with the center of the trap where the intensity is maximum, resulting in absorptive heating and optical damage.

In order to overcome the disadvantages of Gaussian optical trap, several alternative optical beam patterns were proposed. One such optical beams is the Laguerre-Gaussian (LG) beam [19]. The LG beams sometimes also referred as the vortex beam, has an advantageous intensity profile, where the center of the optical trap has an irradiance value of 0. Unlike conventional trap, the forces which construct the trap are not gradient forces but the angular momentum in a swirling optical field and the particle gets trapped at the center. Diffractive optical elements such as spiral phase plate [20] and forked grating [21] are used for generation of the vortex beams. In this direction of research, different DOEs with interesting and useful intensity profiles such as flower shaped beams [22], higher order Bessel beams [23], asymmetric vortex beams [24], beams with structured orbital angular momentum [25] and focused vortex beams [26] were designed. While the above intensity patterns have many advantages, in most of the cases, the design and fabrication of DOEs is a challenge and the efficiency is poor due to combining different phase functions [27]. In [24, 26], the DOEs are not phase-only but use both amplitude and phase modulation simultaneously which results in a poor efficiency. In [22], the DOE consists of three phase functions which again reduces the effective diffraction efficiency.

Recently, a novel DOE called chiral Fresnel zone plate [28] was designed by implementing zone rotation principle [29,30] on a square Fresnel zone plate [31]. This element generates a twisted intensity and phase distribution and was found to possess optical trapping capabilities. The intensity distribution generated by the DOE is peculiar with a hybrid pattern: Gaussian intensity at the centre and twisted side lobes. Even though the hybrid optical trap configuration may offer a stronger and stable trap, the central spot may heat up the trapped particles and cause optical damage. In this study, we introduce a DOE based on binary square axicon and with zone rotation principle called chiral binary square axicon (CBSA). The focusing characteristics of axicon, in general, are unique with a large focal depth and non-varying intensity pattern [32]. Moreover, the far-field intensity pattern does not have a central Gaussian peak as with Fresnel zone plates. Therefore, it is interesting to investigate the focusing properties of a chiral version of the axicon and its optical trapping capabilities. The beam characteristics of the light modulated by a square axicon and lens have been reported earlier for THz radiation [33].

The paper consists of four sections. In the second section, the design and simulation of the DOE and the calculation of the optical fields using scalar diffraction formulation is presented. Experimental analysis of the CBSA and the optical trapping experiment are discussed in the third section. The results are discussed in the final section.

## 2. Methodology

An axicon, or conical prism, is an optical element which can generate a Bessel-like optical field within its depth of focus and a ring pattern in the far-field [34]. Axicon has various advantages over conventional lenses such as large depth of focus [35] also known as line focus. Besides, manufacturing of diffractive axicons is relatively easier as they have a uniform period unlike diffractive Fresnel lenses whose periods decrease away from the center. The line focus characteristic of axicons has been utilized for beam shaping applications. By combining the optical function of an axicon with a second optical function, the line focus of axicon can be employed as a beam delivery path for the second optical function. One such case is the combining of axicon and a spiral phase plate to generate higher order Bessel beams for optical manipulation [22, 36]. Furthermore, combining an axicon with a varying phase was used for generation of accelerating airy beams [37]. In [4,5,38] the far-field ring pattern of Bessel beam is used for different applications.

In this study, a vertex controllable chirality principle is employed [28] i.e., the number of vertices in the 2D phase of the DOE will yield an optical field with chirality of the same order. In this case, a chirality order $p = 4$, is considered which corresponds to a square axicon. The size of each zone is given as $W_n = n\Lambda/2$, where $\Lambda$ is the period of the CBSA and $n = 0, ½, 1,…$

The zones are rotated linearly with respect to one another as $\theta = Km$, where, $m = 2n$ and $K$ is the angular step size in degrees. A grating period of $\Lambda=320$ μm and a central wavelength of 633 nm are selected for evaluation of the CBSA characteristics. The optical fields generated by the CBSA are studied by combining a quadratic phase function $Q(-1/f) = \exp[(-j\pi/\lambda f)r^2]$, where $r^2=(x^2+y^2)$ and $f = 25$ cm. The QPM and CBSA are combined using modulo-$2\pi$ phase addition technique expressed as $\Phi_{DOE}=[\Phi_{CBSA}+\Phi_{Q(-1/f)}]_{2\pi}$. The CBSA is designed with a binary phase values of 0 and $\pi$ and with an efficiency of 40% [6]. The modulo-$2\pi$ phase addition of CBSA($K=0$) and $Q(-1/f)$ is shown in Fig. 1(a)-1(e). A continuous phase profile of QPM has a theoretical efficiency of 100% and so the effective efficiency is 0.4(CBSA) × 1 (QPM) = 0.4. However, it is a difficult task to fabricate, using lithography procedures, a DOE with a resist profile as shown in Fig. 1(e) [6]. In the case of fabrication, a binary version (0, $\pi$) of QPM can be combined with the CBSA, where the modulo-$2\pi$ phase addition technique is reduced to a simple logical-OR operation [26], resulting in a binary phase profile. In this way, the complexity in fabrication can be reduced at the expense of a decrease in the diffraction efficiency. The reduced efficiency in turn can be compensated using a higher input optical power. In the current manuscript, the demonstration is carried out using phase-only spatial light modulator (SLM) and therefore QPM with a continuous phase profile is used to obtain a higher efficiency.

For analysis, CBSA with $K=0°$ to $10°$ in steps of $2°$ is considered. The phase profiles of CBSA before and after modulo-$2\pi$ phase addition of $Q(-1/f)$ are shown in the column – 1 and column – 2 of Fig. 2, respectively. Assuming that the light incident on the DOE is a plane wave with a uniform intensity profile, the amplitude and phase profile at any axial plane is calculated using Eq. (1) as

$$E(z) = \left[\exp\left(j\left[\Phi_{CBSA}+\Phi_{Q(-1/f)}\right]_{2\pi}\right)\right] \otimes Q(1/z), \quad (1)$$

where the symbol '$\otimes$' represents a 2D convolution operator.

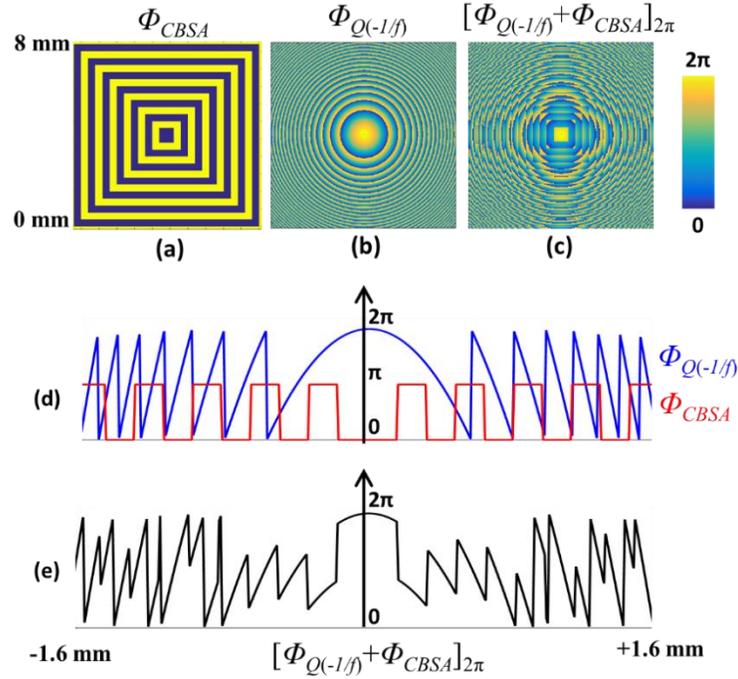

Fig. 1. Phase masks of (a) CBSA (b) $Q(-1/f)$ and (c) $[\Phi_{CBSA}+\Phi_{Q(-1/f)}]_{2\pi}$. (d) Phase profiles of central part of CBSA and $Q(-1/f)$ and (e) phase profile of the central part of $[\Phi_{CBSA}+\Phi_{Q(-1/f)}]_{2\pi}$.

The intensity and phase images generated by the CBSA at the focal plane of $Q(-1/f)$ for different values of $K$ are shown in the third and the fourth columns of Fig. 2, respectively. With an increase in the value of $K$, the twist in the intensity and phase images are increased and there is a redistribution of intensity from the side lobes to the center. For $K=0°$, there is no redistribution of light to the center. For $K=2°$, $4°$, $6°$, $8°$ and $10°$ there is a redistribution of 0.3%, 0.7%, 1%, 1.2% and 1.3% of the light intensity to the center, respectively. Besides, a closer look at the magnified images of the center of the intensity pattern, for $K=2°$, $4°$, $6°$, $8°$ and $10°$ in Fig. 2 reveals that the light is redistributed to eight intensity peaks located around the optical axis. The case $K=4°$ has a zero central maximum, a beam pattern suitable for optical trapping of absorptive samples. The inner radius of the intensity pattern for $K=4°$ is roughly one-fifth of the spacing between the intensity peaks for $K=0°$. The normalized intensity profile at $y=0$ is plotted for different values of $K$ in the fifth column of Fig. 2. The intensity plot shows once again that the case $K=4°$ has an intensity profile suitable for trapping absorptive samples. The beams generated by the DOEs were studied for any evidence of OAM by plotting the Poynting vectors which in the case of scalar optical fields are calculated as the product of the gradient of the phase and intensity distribution of the beam [39]. The Poynting vector plot for $K=4°$ with a contour plot of the intensity values in the background are shown in Fig. 3. Even though at the first glance, the Poynting vector plot for $K=4°$ seems to have a circulating field about the eight points, a closer look reveals that the fields between consecutive points are not propagating in one direction which indicates the absence of OAM. A similar plot obtained for all the other values of $K$ showed that the beams do not possess any OAM. The intensity and phase profiles were simulated along the longitudinal direction by varying $z$ in equation (1) and it was found that the intensity pattern rotated about the optical axis in all the cases of $K>0°$, while no intensity rotation was noticed in the case of $K=0°$. The video of intensity variation when the $z$ value was varied from 23 cm to 25 cm for $K=0°$ and $K=4°$ are shown as visualizations – 1 and 2 respectively, in Fig. 2. Similar intensity rotations were noticed for other values of $K>0°$. The longitudinal intensity profiles are shown for $K=0°$ and $K=4°$ in Figs. 4(a) and 4(b) respectively. Intensity rotation as above has been reported earlier by Dudley, *et. al*,. when higher order Bessel beams were generated, using ring slits with opposite topological charges, and superimposed [40,41]. They have reported a similar observation of rotation of intensity pattern around the optical axis while there was no global OAM. Moreover, the number of petals generated by the superposition was twice that of the topological charge. In this case, the number of peaks surrounding the optical axis is twice that of the number of vertices of the DOE. We believe that in the case of CBSA, even though the global OAM is zero, there is a superposition of vortex states which causes the rotation of the intensity pattern around the optical axis when the axial distance is varied [40].

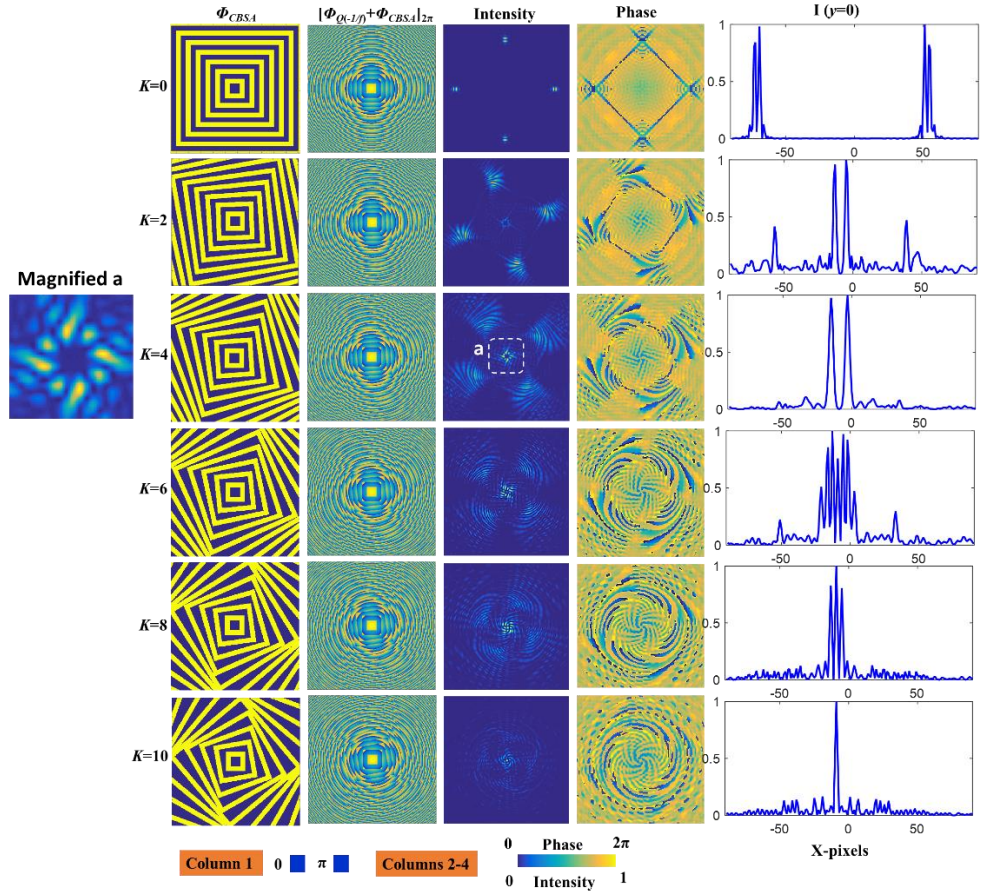

Fig. 2. Column:1 Phase images of CBSA, Column:2 Phase images after modulo-2π phase addition of $Q(-1/f)$, Column:3 Intensity images at the focal plane, Column:4 Phase images at the focal plane for $K = 0°$ to $10°$ in steps of $2°$, Column:5 Intensity plots along $y=0$ Video clips showing the variation of the pattern when z was varied from 23 cm to 27 cm for $K=0°$ and $K=4°$ are given in Visualization – 1 and Visualization – 2 respectively.

The modulo-2π phase addition of the QPM with CBSA has an additional advantage other than bringing the far-field diffraction pattern at the focal plane of the QPM. Combining of QPM with CBSA results in a quasi-achromatic DOE i.e., when there is a change in wavelength, even though the focal length changes, the size of the first order intensity pattern remains a constant. For instance, when λ increases, the focal length of QPM decreases, on the other hand, the first order diffraction spot spacing of axicon increases and they both compensate to yield a wavelength independent behavior. The quasi-achromatic behavior has been extensively studied in [37]. When the wavelength changes, there will be only a slight redistribution of light intensity between the diffraction orders unlike the case without QPM [26].The wavelength was varied from 535 to 735 nm and the variation in the spacing between the horizontal spots or vertical spots are calculated for CBSA for $K=0°$ and $K=4°$ in the presence of QPM. The intensity profile along ($y=0$) is plotted for $\lambda_{a,b,c}$ = 535 nm, 635 nm and 735 nm for $K = 0°$ and $4°$ in Fig (5). It can be noted that even though the focal length was varied as $f_2=f_1\lambda_1/\lambda_2$ [38], where $f_1$ and $f_2$ are focal lengths corresponding to the wavelengths $\lambda_1$ and $\lambda_2$ respectively, the spacing between the intensity spots remained unaltered indicating the quasi-achromatic behavior. The focal length values of QPM $f(\lambda_a,\lambda_b,\lambda_c)$ = 29.7 cm, 25 cm and 21.6 cm.

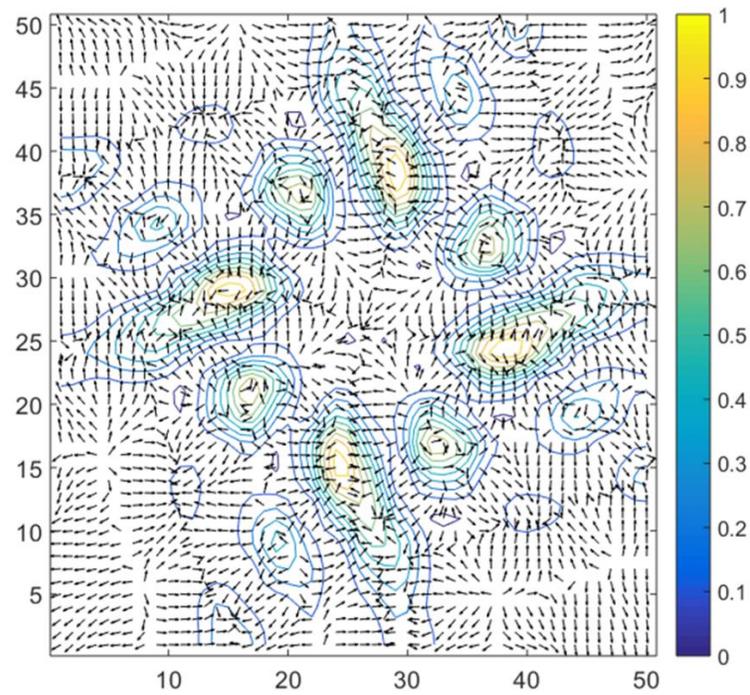

Fig. 3. Map of Poynting vector field plot for $K = 4°$.

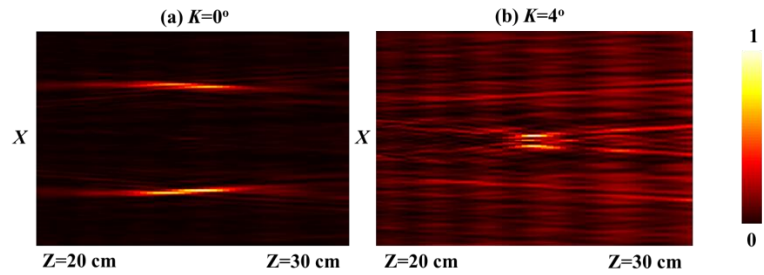

Fig. 4. Longitudinal intensity patterns of CBSA without the quadratic phase mask for (a) $K = 0°$ and (b) $K = 4°$.

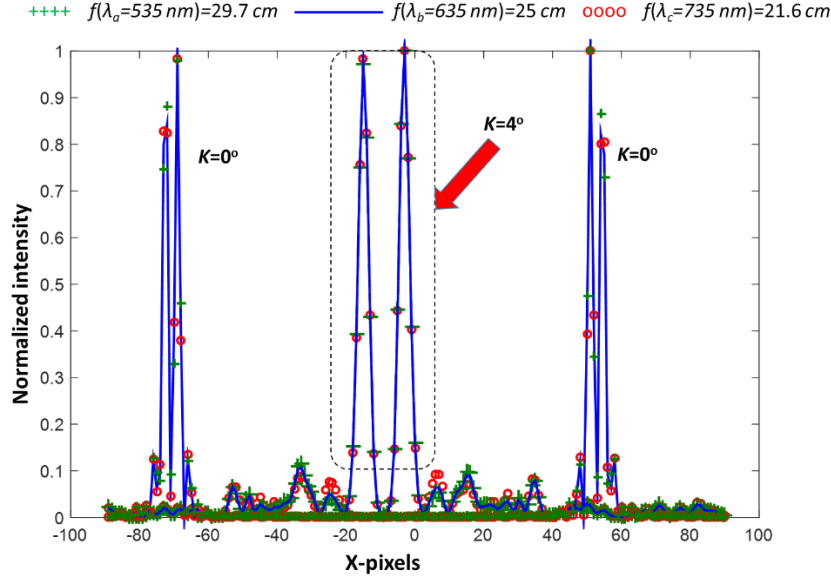

Fig. 5. Plot of the normalized intensity for K=0º (outer) and K=4º(center) for $f(\lambda_a,\lambda_b,\lambda_c)$ = 29.7 cm, 25 cm and 21.6 cm.

From the above analysis, it is found that CBSA generates a twisted intensity and phase profiles, due to the linear rotation of the zones with respect to one another. The intensity redistributes to the center by a small fraction with a space, which is only one-fifth of that of the outer lobes. While the Poynting vector plot revealed the absence of a global OAM, the intensity pattern was found to rotate around the optical axis when the axial distance was varied. Considering the absence of the central peak, the CBSA with $K$=4º was selected for the optical trapping experiment. Similar to [40], there was no intensity rotation observed in the near field but when the axial distance was increased, the rotation was present for all cases $K$>0º.

## 3. Experiments

The intensity patterns generated by the CBSA are experimentally verified using a setup shown in Fig. 6. A He-Ne laser with a wavelength of 632.8 nm was used. The light from the laser is spatially filtered using a pinhole of diameter 80 μm and collimated using a lens $L_1$ with a focal length of $f_{L1}$=18 cm. The collimated light is incident on a phase-only spatial light modulator (Holoeye PLUTO, 1920 × 1080 pixels, 8 μm pixel pitch, phase-only modulation) (SLM) after being polarized using a polarizer along the active axis of the SLM to create full modulation. On the SLM, $\Phi_{DOE}=[\Phi_{CBSA}+\Phi_{Q(-1/f)}]2\pi$ was displayed for different values of $K$ with a period of $\Lambda$=320 μm and focal length $f$=25 cm. A camera (Thorlabs DCC3260C, 1936 × 1216, 5.85 μm pixel pitch) is mounted at a distance of $z$ = 25 cm from the SLM. The uniform intensity assumption in the simulation is satisfied by allowing the light to expand after spatial filtering and installing an iris at the entrance of the SLM to allow an approximately uniform intensity. The intensity profiles for $K$ = 0º to 10º in steps of 2º for $f$ = 23 cm – 27 cm in steps of 1 cm is shown in Fig. 7. The intensity patterns at $f$ = 25 cm matches with the simulation results shown in Fig. 2. The intensity patterns when $f$ is varied from $f$-3 cm to $f$+3 cm shows the rotation of the intensity profile around the optical axis. The phase pattern of CBSA for $K$=4º is displayed on the SLM and preliminary optical trapping studies were conducted. The details of the optical trapping experiment are given in the Appendix A.

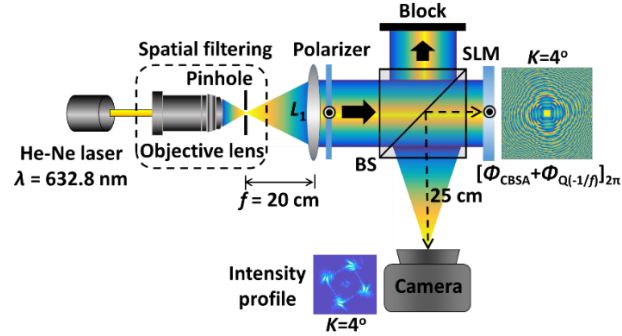

Fig.6. Experimental setup used for evaluating the far field diffraction patterns of CBSAs.

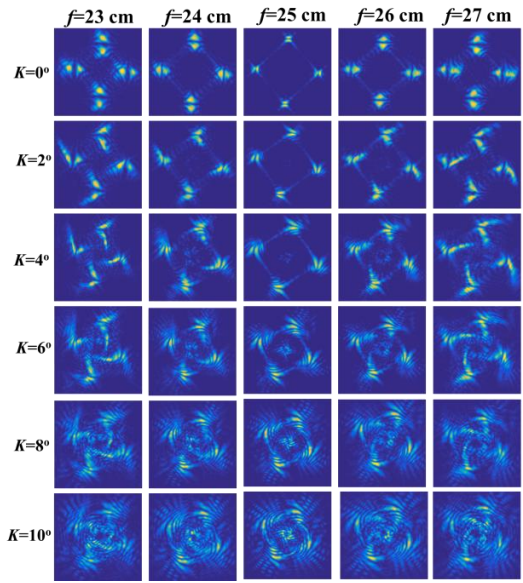

Fig. 7. Experimental intensity patterns recorded at a distance of $z = 25$ cm from the SLM for $K = 0º – 8º$ in steps of $2º$ and $f = 23$ to $27$ cm.

## 4. Summary and conclusions

In conclusion, we have introduced a novel DOE called chiral binary square axicon (CBSA) designed by a linear rotation of the half period zones with respect to one another. The phase profile of CBSA is combined with a QPM to bring the far-field diffraction pattern of CBSA at the focal plane of the QPM. The rotation of the zones in turn twists the intensity and phase patterns at the focal plane of the QPM. The optical fields generated by the DOE are studied using scalar diffraction formulation and the presence of whirlpool-like phase and intensity profiles are confirmed. Poynting vector study revealed that there is no global OAM in the optical field. However, the intensity patterns were found to rotate around the optical axis along the longitudinal direction. Earlier studies on superposition of higher order Bessel beams generated such rotating intensity patterns [40]. In those cases as well there was no global OAM but superposition of vortex states. The designed DOE is not a classical structure and therefore the analysis is not direct as in the case of ring slits. We believe that further theoretical investigations are necessary to completely understand the behavior of CBSA.

The integration of QPM with CBSA introduces quasi-achromatic properties resulting in a high stability for changes in wavelength. As a result, the DOE designed for a particular wavelength can also be used for other wavelengths. The DOE is experimentally evaluated using an optical setup with a phase-only SLM and the intensity images at the focal plane of the QPM were recorded. The experimental results matched well with the simulation results. Preliminary studies were carried out using a regular optical trapping setup with two laser sources and a phase-only SLM. When the intensity patterns generated by the CBSA is focused on a trap specimen, the specimen rotated around the center of the optical trap.

In this study, the simulation, experimental and preliminary optical trapping results confirm the suitability of CBSA for optical trapping applications. CBSA has many advantages compared to existing DOEs which are used for optical trapping applications. CBSA is a binary DOE and has an uniform period which makes the fabrication procedure much easier for manufacturing compared to other DOEs such as spiral phase plates [20], chiral Fresnel zone plates [28], hybrid DOEs [22, 26, 27], etc., Secondly, the absence of central peak in the optical field expands the applicability of CBSA to highly absorptive optical trapping specimens.

In the current manuscript, only a chirality order of 4 is studied. Preliminary studies on different chirality orders have been reported using Fresnel zone antennas designed for microwave [43] and optics [44]. We believe that further studies are necessary with other chirality orders in order to understand the optical trap qualities better. Advanced studies with the optical trapping experiment is required to compare the optical trapping properties with other DOEs.

### Acknowledgements

We thank Dr. Vijay Kumar, Indian Institute of Science Education and Research, Pune, India and Prof. Andrew Forbes, WITS University, South Africa, for the useful discussions. The work by IVM was carried out within the framework of Tomsk Polytechnic University Competitiveness Enhancement Program.

### Funding

A.Vijayakumar, Mani Ratnam Rai and Joseph Rosen were supported by the Israel Science Foundation (ISF) (Grants No. 1669/16) and by the Israel Ministry of Science and Technology (MOST). B. Vinoth and Chau-Jern Cheng were supported by Ministry of Science and Technology, Taiwan.

**Appendix A**

The schematic of the optical tweezer setup is shown in Fig. A.1. The setup consists of two laser sources: a high power (5W) laser with a center wavelength $\lambda_C = 532$ nm (green) for optical trapping and another He-Ne laser emitting at $\lambda_C = 632.8$ nm (red) is used for imaging the optical trapping procedure [42]. The optical power at the center of the trap in the case of CBSA with $K=4°$ was approximately 10 mW. The spatially filtered and collimated beam of green laser source is incident on a reflective phase only SLM (Jasper Display Corp: pixel number: 1920 ×1024; pixel pitch: 6.4 μm), a 4F system is embedded after the SLM and the beam is focused by a 100X objective lens (NA=1.4) for the trapping. We have used live *Candida rugosa* (ATCC® 200555™) as a specimen (commonly known as yeast) and the trapping is recorded in real time by an imaging system using the He-Ne laser. The He-Ne laser of the imaging system is also spatially filtered and passed through the 4F imaging system and the image of the trap is recorded by a CMOS camera (Thorlabs: 1024 × 1280; pixel pitch: 5.2 μm). The phase pattern of CBSA for $K=4°$ are displayed on the SLM and the trapping studies were conducted.

In the first experiment, a constant phase value was displayed on all the pixels of the SLM and the optical trapping was carried out using a regular Gaussian intensity profile. The phase pattern of the CBSA with $K=4°$ is displayed on the SLM and the beam generated at its focal plane was reimaged using the 4F optical system. The images of the yeast specimen sample before and after trapping using the beam pattern for $K = 4°$ is shown in Fig. A.2. When the yeast specimen was overlapped with the trap beam, the yeast sample was trapped as well as rotated. However, when the phase pattern of CBSA for $K=0°$ is displayed, no such rotation was noticed for any case. The trap video is given in Video – 3 for $K=4$. We believe further studies are required in order to understand the intensity rotation characteristics of the CBSA and may be used for controlled rotation optical trapping experiments in the future.

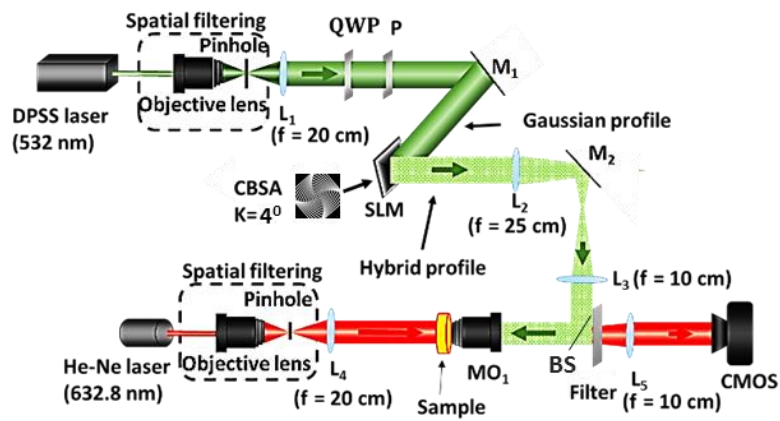

Fig. A.1 Schematic of the optical tweezer set up.

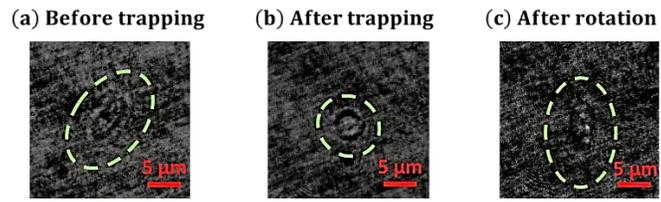

Fig. A.2 Microscopic images of *Candida rugosa* (a) before trapping, (b) after trapping and (c) after rotation for $K = 4°$.